\documentclass[conf]{new-aiaa}
\usepackage[utf8]{inputenc}
\usepackage{comment}
\usepackage{graphicx}
\usepackage{amsmath}
\usepackage[version=4]{mhchem}
\usepackage{siunitx}
\usepackage{longtable,tabularx}
\newcommand{\R}{\mathbb{R}}
\usepackage[ruled,vlined]{algorithm2e}
\usepackage{algorithmic}
\usepackage{comment}
\usepackage{subfigure}
\usepackage{makecell}
\title{Bifurcation-Based Guidance Law for Powered Descent Landing}

\author{Neon Srinivasu
\footnote{Department of Mechanical and Aerospace Engineering, Syracuse University, Syracuse, NY 13244, USA,  neonsrin@syr.edu}, Amit Shivam \footnote{Research Assistant, Department of Electrical and Computer Engineering, University of Porto, Porto, Portugal, amitshivam@alum.iisc.ac.in}, and Nobin Paul \footnote{Post Doctoral Researcher, Department of Aerospace Engineering, KAIST, nobinpaul@kaist.ac.kr}}

\begin{document}
	\footnotetext{The paper was accepted for presentation at the AIAA SciTech Forum 2026 but was later withdrawn by the authors -  budget constraints.} 
\maketitle

\begin{abstract}
This paper develops a new guidance law for powered descent landing of a rocket-powered vehicle. The proposed law derives the acceleration command for a point mass model of the vehicle by expressing velocity  as a dynamical system undergoing supercritical transcritical bifurcation with three bifurcation parameters. The parameters are designed such that the stable equilibrium points of the velocity dynamics correspond to the guided targeting state, that is, the landing point. Numerical simulations are performed to demonstrate the working of the proposed guidance law.  
\end{abstract}

\section{Introduction}
\lettrine{P}{owered} descent phase is arguably the most intricate part of the entry, descent, and landing part of the precision landing technology. Powered Descent Guidance(PDG) law primarily determines the landing vehicles direction of motion and speed leading it to fly a trajectory satisfying targeting conditions, such as to reach the designated landing site. In this regard, an utmost important requirement by the PDG law is that it has to be computationally tractable on physical flight processors. \\

 PDG algorithms can be broadly classified into three categories: Polynomial guidance laws ~\cite{1,2,3,4,5,6,8,7}, optimal control ~\cite{13,9,ross,10,10r,10rr,11,12,n5,int1,n0,n1,n2,n3,n4,n6,n7} and learning based methods ~\cite{l1,l2,l3}. Its been over 50 years since the famous Apollo Powered Descent Guidance(APDG) law ~\cite{1} landed the Apollo Lunar Module on the Moon. Owing to its computational simplicity and closed-form solutions, APDG law has been the baseline powered descent guidance approach for many actual missions ~\cite{2,3,4,5}. APDG and Apollo E-guidance law ~\cite{6} assumes a thrust acceleration vector profile as quadratic and linear functions of time, respectively. For APDG law the coefficients of the quadratic polynomial are determined to satisfy initial and final positions and velocity constraints along with final acceleration. In E-guidance law only initial and final positions and velocity constraints can be satisfied. The main issue of polynomial guidance  laws is the determination of time-to-go that specifies the burn time. Time-to-go  determines the fuel consumed and whether the  thrust will saturate between minimum and maximum allowable bounds. When APDG law is used on actual missions the time-to-go is determined using trial and error method on ground. In Ref.~\cite{8} acceleration is expressed as functions of position and velocity with two tunable gains and specific choices of gains leads to APDG and E-guidance laws. The tuning parameters are used as trade off quantity between trajectory shaping and fuel usage. Ref.~\cite{7}  recently investigated  on a family of guidance laws where acceleration is expressed as fractional polynomial of time-to-go and showed that APDG and E-guidance laws belong to that family.   \\

Optimal control based PDG laws are widely investigated in the literature because fuel optimality is a necessary requirement for landing missions. Fuel optimal PDG problem can be solved using direct and indirect methods. Direct methods involve approximating the optimal control problem with collocation methods  and solving it numerically using nonlinear optimization methods. Indirect method concerns with Pontryagin's maximum principle to form Hamiltonian and first-order necessary conditions to solve two point boundary value problem. Interested readers are directed to Ref.~\cite{13} for more details on optimal control theory.  Ref.~\cite{9} first used indirect method to formulate guidance law which minimizes commanded acceleration and the obtained result was surprisingly E-guidance law. In that work, cost function is modelled as integral of quadratic acceleration, which is not theoretically fuel optimal but control effort optimal. Fuel optimal consideration requires integral of thrust magnitude not quadratic index ~\cite{ross}. An important result optimal control method reveals is that the solutions to fuel optimal PDG are such that thrust profile has bang-bang structure operating at either maximum or minimum thrust bounds. In a constant gravity field,  three dimensional fuel optimal PDG problem Refs.~\cite{10,10r,10rr} showed that there can be at most two switching in optimal thrust magnitude. In non-constant gravity field, depending on how gravity is modelled  number of thrust switching varies. Optimal Thrust profile with thrust pointing as a constraint in PDG problem is analyzed in Refs.~\cite{11,12} \\

Although, direct methods give enough information on how thrust should behave for fuel optimality, when several operational constraints on the generated trajectory are considered, such as, velocity, thrust pointing, glide-slope it becomes difficult to analyze the problem analytically. Thus several  numerical optimization techniques, are investigated in Refs.~\cite{n5,int1,n0,n1,n2,n3,n4,n6,n7}.  Among the optimization methods, convex optimization  is frequently used to solve PDG problem where, convex problem is formulated and solved using interior point method ~\cite{n5,int1}. Since, thrust is non-convex input constraint (due to non-zero lower thrust bound), it can be mapped to a convex set by introducing additional slack variable and can  be solved using lossless convexification method ~\cite{n0,n1,n2,n3,n4,n6,n7}. That method has flown onboard a real rocket dubbed G-fold algorithm ~\cite{n6,n7}. Main drawback of optimization methods is that  is computationally expensive and may not be feasible to implement onboard the vehicle. Moreover, convergence of direct methods relies heavily on initial guess of the unknown variables and global convergence to a local optimum isn't guaranteed. If the optimal solution's convergence time  is greater than the guidance cycle then the system fails. \\

In recent times, learning based methods have been used to find solution to fuel optimal PDG problem. Ref.~\cite{l1} applied deep neural network to learn relationship between initial condition and optimal state-control pairs. Therein, calculating all of the state-control variables demands heavy computation. Ref.~\cite{l2} combined the supervised learning and optimal control theory to reduce the dimension of learning space thereby reducing computation load. Therein, the learning process is supported by necessary conditions on Pontryagins's minimum principle to find critical parameters required to obtain optimal solution. Ref.~\cite{l3} extends the work in Ref.~\cite{l2} by using mixed-input deep neural network to map the relationship between the optimal problem inputs and the critical parameters, which reduces the computational time. In general, to achieve high accuracy of the learning result it is required to generate large-scale neural networks to calculate optimal state-control pair for given input demanding heavy computational load. Also, learning based methods lacks theoretical proof of convergence to optimal solution.   \\

This work proposes a new bifurcation based PDG law (BPDG) based on bifurcation phenomenon in dynamical systems theory \cite{anjaly2020target,2022standoff,srinivasu2022guidance}. The velocity dynamics is modelled as a dynamical system undergoing supercritical transcritical bifurcation with three bifurcation parameters. Stability properties of the proposed system is analyzed and closed-form analytical expressions are derived for determining the bifurcation parameters as a function of initial position and velocity and final targeting condition. Further, deterministic performance characteristics of the position, velocity and acceleration profiles, that is, the monotonic nature along with closed-form expression for the extreme values is established.  
Overall, BPDG law enjoys same computational simplicity and elegancy of APDG. \\

The remainder of this paper is arranged as follows: Section \ref{sec1} defines
the problem statement. The proposed velocity dynamics 
is discussed in Sec.~\ref{sec2}. Section~\ref{sec3} presents proposed guidance law,
which is followed by  simulation
studies  in Sec.~\ref{sec4}.  Section~\ref{sec5} concludes this paper.

\section{Problem Statement} \label{sec1}
\begin{figure}[h!]
				\centering
				{\includegraphics[scale=0.9]{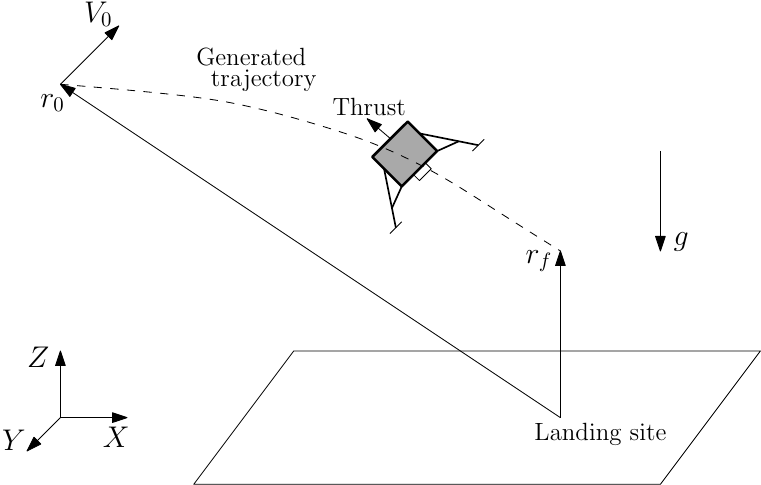}}
				\label{}
				\caption{{Problem Description}}
				\label{fig.1}
			\end{figure}
   In cartesian coordinate system the three-dimensional point-mass powered descent dynamics for planetary landing is described by
   \begin{eqnarray}
						\dot{r}&=&V   \\
						\dot{V}&=&g+A\\
						A&=&\frac{\textup{T}(t)}{m(t)}\\
						\dot{m}&=&- \frac{\lVert \textup{T}(t)  \rVert}{ v_\textup{ex}}		 
					\end{eqnarray}
   where $r,V,A, T \in \R^3$ are the position, velocity, acceleration, and thrust vector of the vehicle and $g$ is constant gravitational acceleration vector. $v_\textup{ex}$ is the exhaust velocity of the engine and is a constant. The aerodynamic forces on the vehicle are neglected by assuming the planet atmosphere is sufficiently thin. Subscripts $\textup{0}$ and $\textup{f}$ denote initial and final values of corresponding physical quantities. The guidance objective is to determine $A$ that transfers the vehicle from the powered descent initiation(PDI) position $r_0$ to the  powered descent termination(PDT) position $r_\textup{f}$. It is desired that the guidance command $A$ has deterministic performance characteristics and easily computable.

\section{Proposed Velocity Dynamics} \label{sec2}
\begin{figure}[h!] 
				\centering
				{\includegraphics[scale=1]{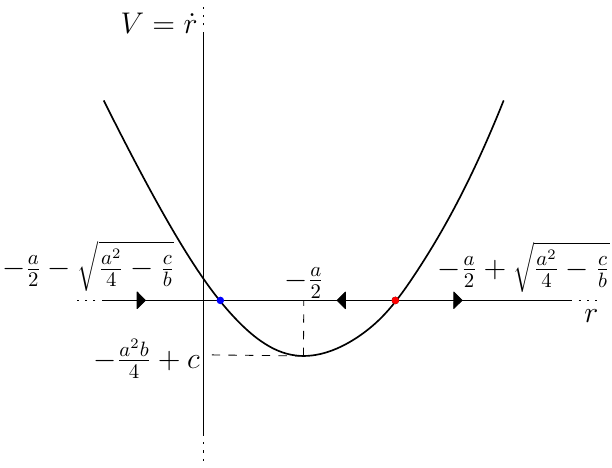}}
				\label{}
				\caption{{Velocity profile}}
				\label{figo1}
			\end{figure}

\begin{figure}[h!] \label{fig2}
	\centering
	\subfigure[{Bifurcation diagram for $c<0$}]{
		\includegraphics[width=8cm]{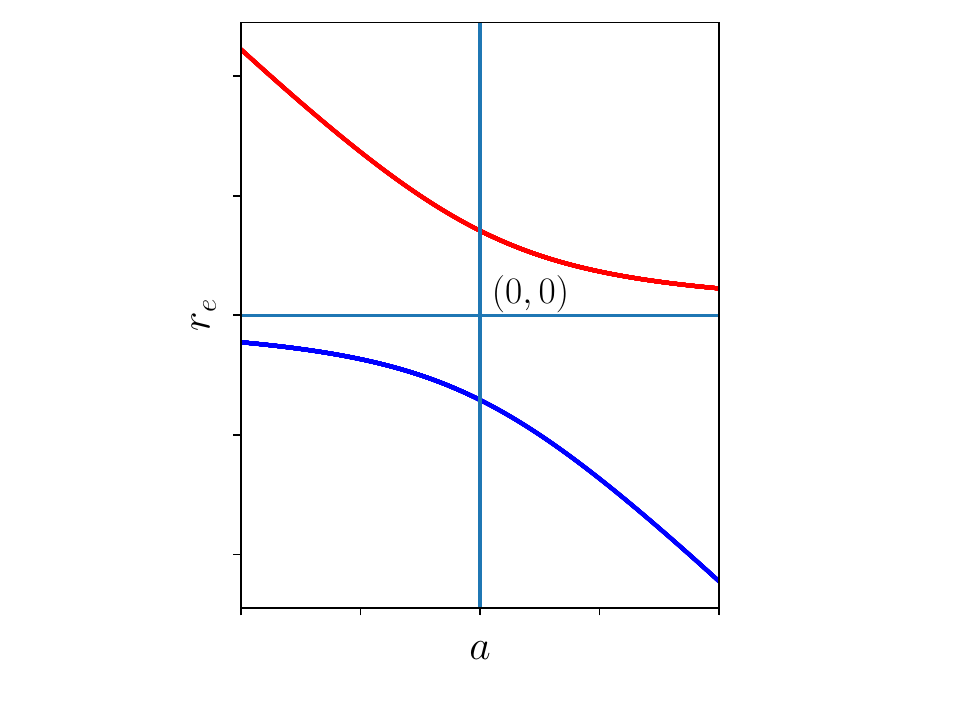}}
	\label{fig6.1}
	\subfigure[{Bifurcation diagram for $c=0$}]{
		\includegraphics[width=8cm]{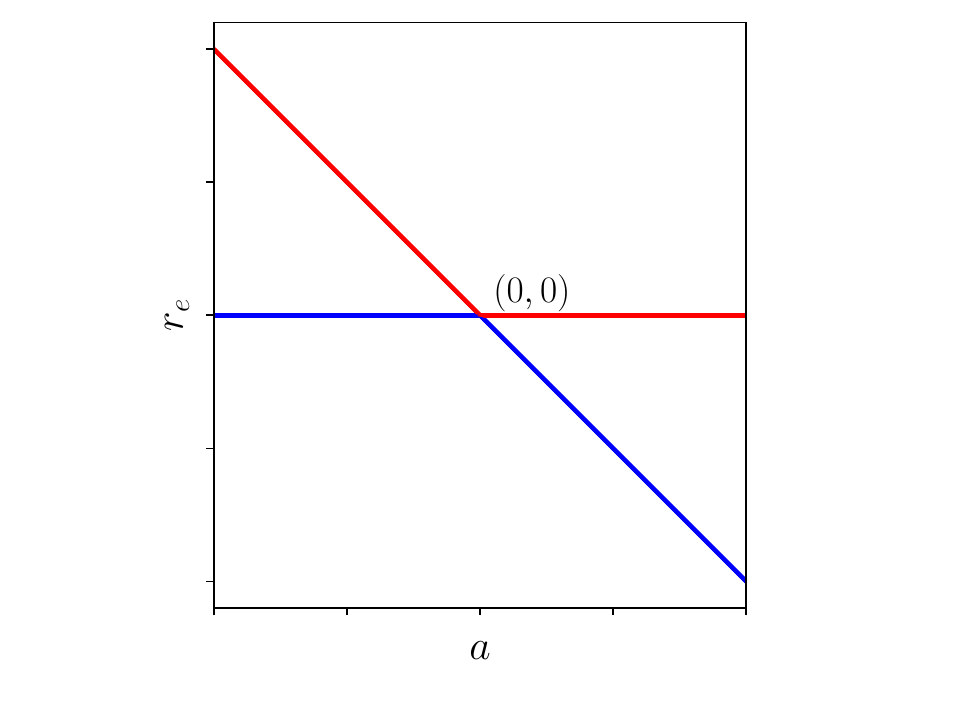}
	}
	\label{fig6.2}
 	\subfigure[{Bifurcation diagram for $c>0$}]{
		\includegraphics[width=8cm]{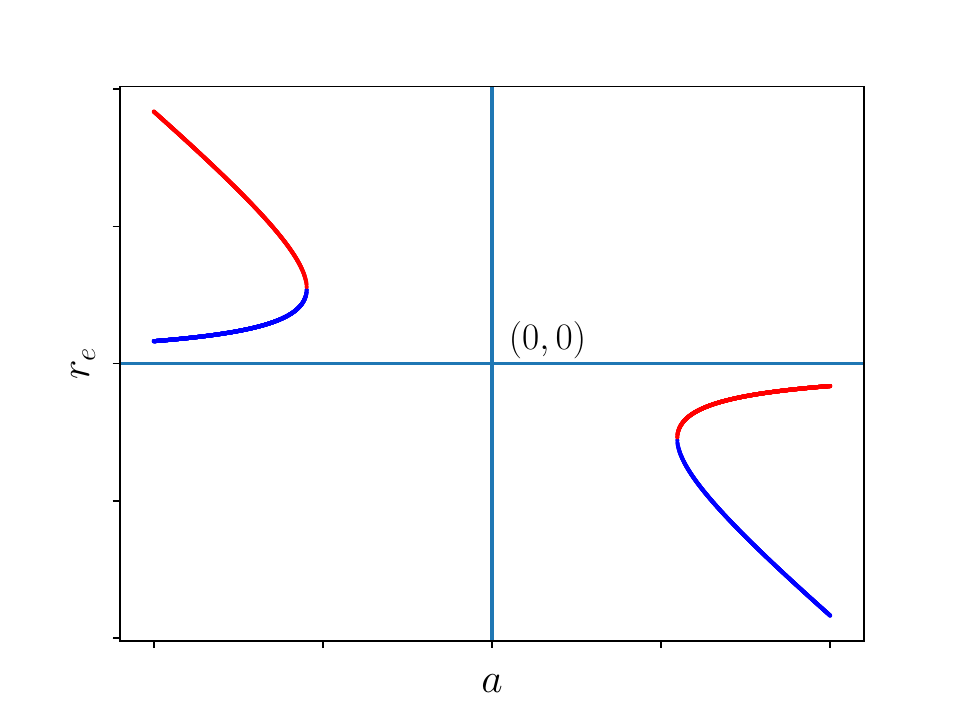}
	}
 \caption{Bifurcation diagrams for different choices of $c$}
	\label{fig6}
\end{figure}

The velocity dynamics of the landing vehicle in each of the three axis is proposed as, 
\begin{equation} \label{aa}
    V_{\textup{i}}=\dot r_{\textup{i}}=\left(a_{\textup{i}}r_{\textup{i}}+r_{\textup{i}}^2\right) b_{\textup{i}}+c_{\textup{i}
    }, \hspace{0.25cm} {\textup{i}}={x,y,z}
\end{equation}
where $a,b,c \in \R$ are bifurcation parameters to be designed. Subscript $\textup{i}$ corresponding to three axes are excluded in few figures and analysis and, all the results are valid true for all  $\textup{i}={x,y,z}$. The equilibrium points $r_{\textup{ie1}},r_{\textup{ie2}}$ of the considered dynamical system are obtained using $\dot V=0$, 
   \begin{eqnarray} \label{eq1}
						r_{\textup{ie1}}&=&-\frac{a_\textup{i}}{2}+\sqrt{\frac{a_\textup{i}^2}{4}-\frac{c_\textup{i}}{b_\textup{i}}} \\ 
						r_{\textup{ie2}}&=&-\frac{a_\textup{i}}{2}-\sqrt{\frac{a_\textup{i}^2}{4}-\frac{c_\textup{i}}{b_\textup{i}}} \label{eq2}
   \end{eqnarray}
   Further, the stability properties can be determined using,
	\begin{eqnarray}\label{eqnaa}
		\left.	\frac{\partial \dot r_i}{\partial r_\textup{i}}\right\vert_{r=r_{\textup{ie1}}}&=&2\sqrt{\frac{a_\textup{i}^2b_\textup{i}^2}{4}-c_\textup{i}b_\textup{i}}  \\
		\left.\frac{\partial \dot r_i}{\partial r_\textup{i}} \right\vert _{r=r_{\textup{ie2}}}&=&-2\sqrt{\frac{a_\textup{i}^2b_\textup{i}^2}{4}-c_\textup{i}b_\textup{i}}  \label{eqnab}
	\end{eqnarray}
 Eqs.~(\ref{eqnaa}) and (\ref{eqnab}) indicates that the equilibrium point $r_{\textup{ie1}}$ is unstable and $r_{\textup{ie2}}$ stable.
 A typical plot of the variation of $V$ as a function of $r$ is shown in Fig.~(\ref{figo1}) where red and blue dot represents unstable and stable equilibrium points, respectively. The nature of velocity depends on where the initial position $r_0$ lies. This work considers $0 < r_\textup{i0} \leq -a/2$.  The plot of the variation of equilibrium points as a function of parameter $a$ with fixed $b>0$ for different $c$ is shown in Fig.~(\ref{fig6}) where, the blue and red curve represents stable and unstable equilibrium points, respectively. In Fig.~(\ref{fig6}), when $c\leq 0$ there always exist two equilibrium points, one stable and another unstable. But, when $c>0$ equilibrium points exist iff $(a^2/4-c/b)>0$ which can be readily concluded using Eqs.~(\ref{eq1}) and (\ref{eq2}).   

 Using the method of separation of variables the solution of Eq.~(\ref{aa}) is obtained as
\begin{eqnarray} \label{bb}
  r_\textup{i}(t)&=& \sqrt{\frac{P_\textup{i}}{b_\textup{i}}}\tanh{\left( -\sqrt{P_\textup{i}b_\textup{i}}(t+K_\textup{i})\right)}-\frac{a_\textup{i}}{2}\\
  \textup{where}, K_\textup{i}&=&-\frac{1}{\sqrt{P_\textup{i}b_\textup{i}}}\tanh^{-1}\left(\frac{\sqrt{b_\textup{i}}r_\textup{i0}+a_\textup{i}\sqrt{b_\textup{i}}/2}{\sqrt P_\textup{i}} \right)
\end{eqnarray}
where, $P_\textup{i}=(a_\textup{i}^2b_\textup{i})/4-c_\textup{i}$. The settling value of distance dynamics $r_\textup{i}(t)$  as $t \to \infty$ is
	\begin{equation} \label{cc}
		r_i(t \to\infty)= -\frac{a_\textup{i}}{2}-\sqrt{\frac{a_\textup{i}^2}{4}-\frac{c_\textup{i}}{b_\textup{i}}}
	\end{equation}
corresponding to the stable fixed point which is the guided targeting state, $r_\textup{if}=(r_\textup{xf},r_\textup{yf},r_\textup{zf})$.

\subsection{Designing Bifurcation Parameters}
The three Bifurcation parameters $a_\textup{i},b_\textup{i},c_\textup{i}$ are designed to satisfy vehicles initial position and velocity with final position constraints as described below. Referring to Fig.~(\ref{fig.1}) we impose the constraints so that the velocity monotonically decreases from $V_\textup{i0}$ at PDI to zero at PDT.  
The extreme value of $V$ is obtained as,
\begin{eqnarray}
     \frac{dV_\textup{i}}{dr_\textup{i}}&=&a_\textup{i}b_\textup{i}+2b_\textup{i}r_\textup{i}=0\\
     \implies r_\textup{i}&=&-a_\textup{i}/2\\
     V_\textup{i}|_{r=-a_\textup{i}/2}&=&-\frac{a_\textup{i}^2b_\textup{i}}{4}+c_\textup{i} \label{eqnff}
\end{eqnarray}
Vehicles initial Velocity $V_\textup{i0}$  is equated to the extreme value in Eq.~(\ref{eqnff}) which leads to,
\begin{eqnarray}
        -\frac{a_\textup{i}^2b_\textup{i}}{4}+c_\textup{i}&=&V_\textup{i0}\\
        c_\textup{i}&=&V_\textup{i0}+(a_\textup{i}^2b_\textup{i})/4 \label{11}
\end{eqnarray}
Further, The value of $r$ at which the extreme value occurs is equated to vehicles initial position constraint as, 
\begin{equation}
a_\textup{i}=-2r_\textup{i0}\label{22}
\end{equation}
In addition, if $r_\textup{if}$ is the final targeting constraint then the stable equilibrium point $r_{\textup{ie2}}=r_\textup{if}$ which readily gives,
\begin{eqnarray}
    r_\textup{if}&=&-\frac{a_\textup{i}}{2}-\sqrt{\frac{a_\textup{i}^2}{4}-\frac{c_\textup{i}}{b_\textup{i}}} \\
    b_\textup{i}&=&-V_\textup{i0}/(r_\textup{if}-r_\textup{i0})^2 \label{33}
\end{eqnarray}
Eqs.~(\ref{11}),(\ref{22}) and (\ref{33}) gives the closed-form expression for determining  bifurcation parameters. With the specific design of bifurcation parameters and referring to Fig.~(\ref{fig.1}) it can be readily observed that the maximum velocity of the vehicle occurs at the powered descent initiation and it monotonically reduces to zero.    

\subsection{Powered Descent Termination Time}
As seen from Eq.~(\ref{cc}) that the time taken to reach the equilibrium state is infinite. Hence, we consider error bound as, $r(t_\textup{is})-r(t_\textup{if})=\epsilon_\textup{i}$ and calculate the time taken to reach the bound. Solving,
\begin{equation}
   r(t_\textup{si})-r_\textup{if}=\epsilon_\textup{i}
    \end{equation}
    Substituting for $r(t_\textup{si})$ using Eq.~(\ref{aa}) readily gives
    \begin{equation} \label{hh}
            t_\textup{si}=\frac{1}{2b_\textup{i}\left(r_\textup{if}+ \frac{a_\textup{i}}{2}\right)}\ln{\frac{\epsilon_\textup{i}(r_\textup{i0}+r_\textup{if}+a_\textup{i})}{(2r_\textup{if}+a_\textup{i}+\epsilon_\textup{i})\left(  r_\textup{i0}-r_\textup{if}\right)}}
\end{equation}
Eq.~(\ref{hh}) gives the expression for time taken by the vehicle to reach the final landing site with $\epsilon_\textup{i}$ error. Since, $t_\textup{si}$ in each of the three axes depend on corresponding parameters $a_\textup{i},b_\textup{i},c_\textup{i}$, we express 
\begin{equation} \label{44}
 T\textup{s}=\max \{t_\textup{si}; \textup{i}=x,y,z \}   
\end{equation}
 where $T\textup{s}$ is the powered decent termination time.

\section{Guidance Command} \label{sec3}
Differentiating Eq.~(\ref{aa}) leads to
\begin{equation}\label{kk} 
    A_\textup{i}=\dot V_\textup{i}=\ddot r_\textup{i}=2b_\textup{i}r_\textup{i}^3+3a_\textup{i}b_\textup{i}^2r_\textup{i}^2+\left( a_\textup{i}^2b_\textup{i}^2+2b_\textup{i}c_\textup{i}\right)r_\textup{i}+a_\textup{i}b_\textup{i}c_\textup{i}; \hspace{0.25cm} \textup{i}={x,y,z} 
\end{equation}
 \begin{figure}[h!] 
				\centering
				{\includegraphics[scale=1]{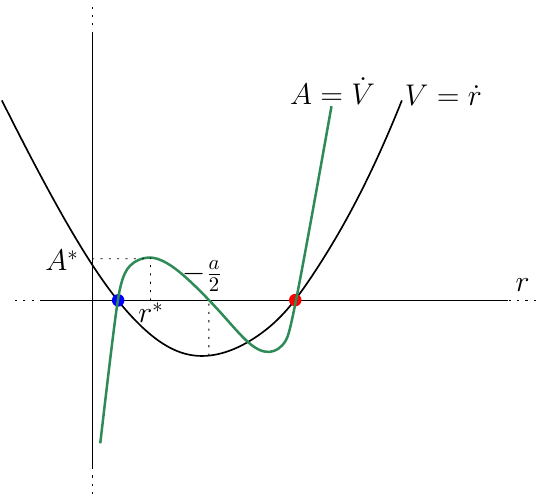}}
				\label{}
				\caption{{Acceleration profile}}
				\label{fig.r}
			\end{figure}
Eq.~(\ref{kk}) describes the acceleration to be commanded to the vehicle so that the velocity and distance dynamics behaves as given in Eqs.~(\ref{aa}) and (\ref{cc}). A typical plot of variation of acceleration experienced by the vehicle with the proposed velocity dynamics in Eqs.~(\ref{aa}) is shown in Fig.~(\ref{fig.r}). The acceleration starts with zero at $r_\textup{i}=-a_\textup{i}/2$ monotonically increasing and  decreasing to zero towards the end of PDT time. The maximum acceleration demanded by the vehicle can be obtained as
\begin{eqnarray}\label{root1}
     \frac{dA_\textup{i}}{dr_\textup{i}}&=&6b_\textup{i}^2r_\textup{i}^2+6a_\textup{i}b_\textup{i}^2r_\textup{i}+(2b_\textup{i}c_\textup{i}+a_\textup{i}^2b_\textup{i}^2)=0\\ 
     \implies r_\textup{i}&=&- \frac{a_\textup{i}}{2}\pm \sqrt{\frac{a_\textup{i}^2}{12}-\frac{c_\textup{i}}{3b_\textup{i}}} \label{root}
     \end{eqnarray}
     Substituting $Q_\textup{i}=\sqrt{a_\textup{i}^2/12-c_\textup{i}/3b_\textup{i}}$ of Eq.~(\ref{root}) in Eq.~(\ref{root1}) leads to
     \begin{equation} \label{acc}
            A_{\textup{i}}^*=-2b_\textup{i}^2Q_\textup{i}^3+\frac{a_\textup{i}^2b_\textup{i}^2Q_\textup{i}}{2}-2Q_\textup{i}b_\textup{i}c_\textup{i}
     \end{equation}

With the analysis discussed so far, the
sequence  to generate the guidance command are detailed in Algorithm~\ref{alg1}. 
 \begin{algorithm}
 	\caption{Guidance logic}
 	\label{alg1}
 	\begin{algorithmic}[1]
 		\STATE   Initialize $r_\textup{i0},V_\textup{i0},r_\textup{fi}$
 		\STATE Determine $a_\textup{i}$ , $b_\textup{i}$, and $c_\textup{i}$  using Eqs.~(\ref{11}), (\ref{22}), and (\ref{33}), respectively 
 		\STATE  Compute $T\textup{s}$ using Eq.~(\ref{44})
 		\STATE Use $(a_\textup{i},b_\textup{i},c_\textup{i})$ to generate the guidance command using  Eq.~(\ref{kk})
 	\end{algorithmic}
 \end{algorithm} 
 
 \section{Simulation Results}\label{sec4}
 Numerical simulations of the BPDG algorithm is demonstrated on a vehicle landing on Mars with martian gravity $g=3.721$ m/s$^2$ . The total initial mass $m_0=m_\textup{V} (\textup{Vehicle mass})+m_\textup{F0} (\textup{Fuel mass})=2000$ kg where,  $m_{\textup{V}}=1500$ kg, and $m_\textup{F0} =500$ kg. The exhaust velocity of the engine $e_{\textup{ex}}=2206.575$ m/s. The allowed error in guided targeting state $\epsilon_\textup{i}=0.1$ m. 
 
 \subsection{Sample Landing Scenario}
 The vehicles initial  conditions at which powered descent starts are tabulated in Table~(\ref{table2})
 \begin{table}[h]
	\centering
	\begin{tabular}{l  c  c }
		\hline \hline
		PDI Position & Velocity & PDT position \\
		\hline 
		$x_0=1900$ m & $V_\textup{x0}=-40$ &$r_\textup{xf}=0$\\
		$y_0=1000$ m &  $V_\textup{y0}=-10$&$r_\textup{yf}=0$\\
		$z_0=3100$ m & $V_\textup{z0}=-50$ & $r_\textup{zf}=5$\\
		\hline \hline		
	\end{tabular} \\ 
	\caption{Initial and Terminal conditions}
	\label{table2} 
\end{table}

\begin{figure}[h!] 
				\centering
				{\includegraphics[scale=0.6]{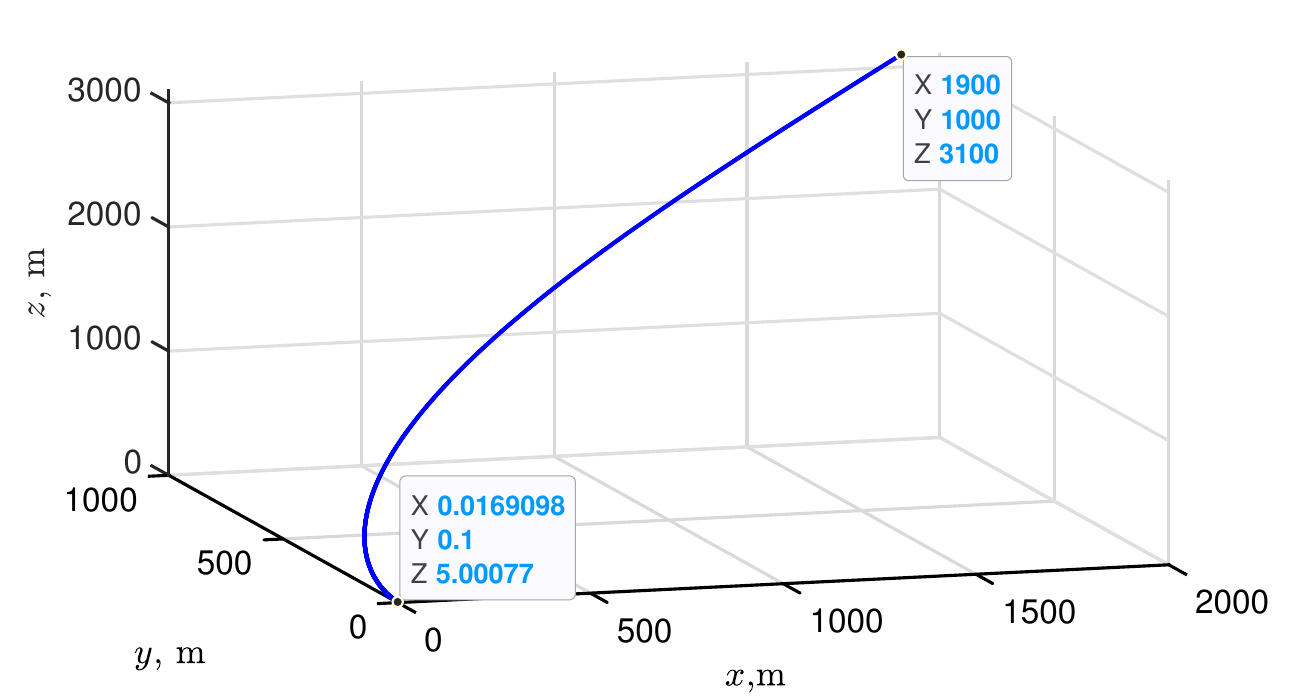}}
				\label{}
				\caption{{Powered Descent Trajectory}}
				\label{figa}
			\end{figure}

\begin{figure}[h!] 
	\centering
	\subfigure[{Position profiles}]
	{\includegraphics[width=7.5cm]{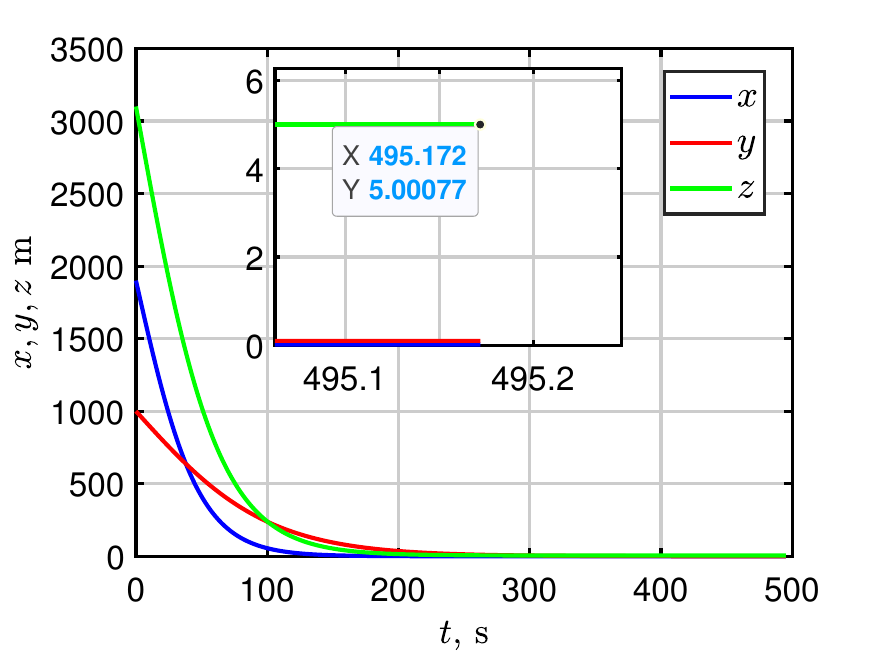}}
	\label{figk1}
	\subfigure[{Velocity profiles}]
	{\includegraphics[width=7.5cm]{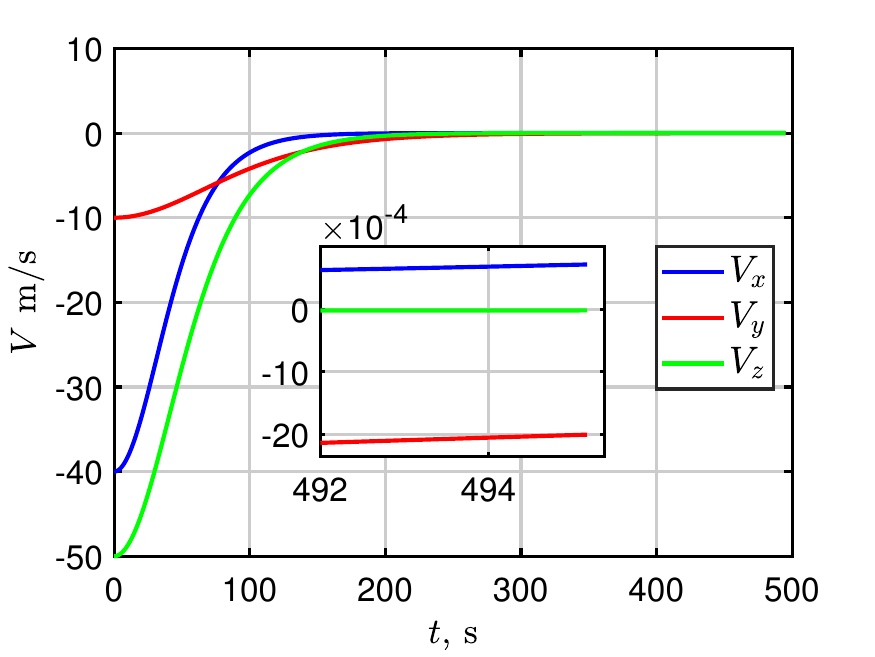}}
	\label{figk2}
	\subfigure[{Acceleration profiles}]
	{\includegraphics[width=7.5cm]{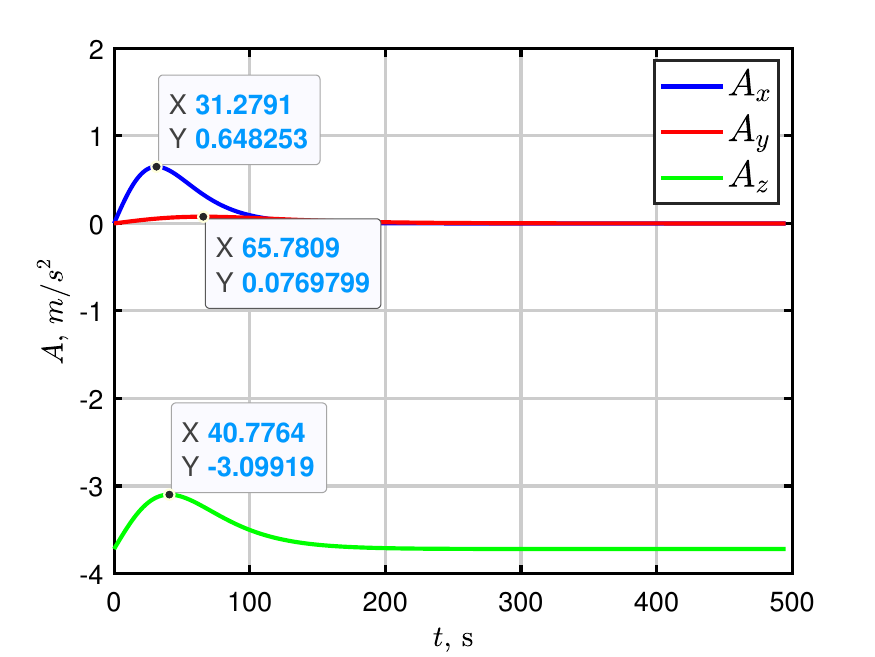}}
	\label{figk6}
	\subfigure[{Fuel mass history}]{
		\includegraphics[width=7.5cm]{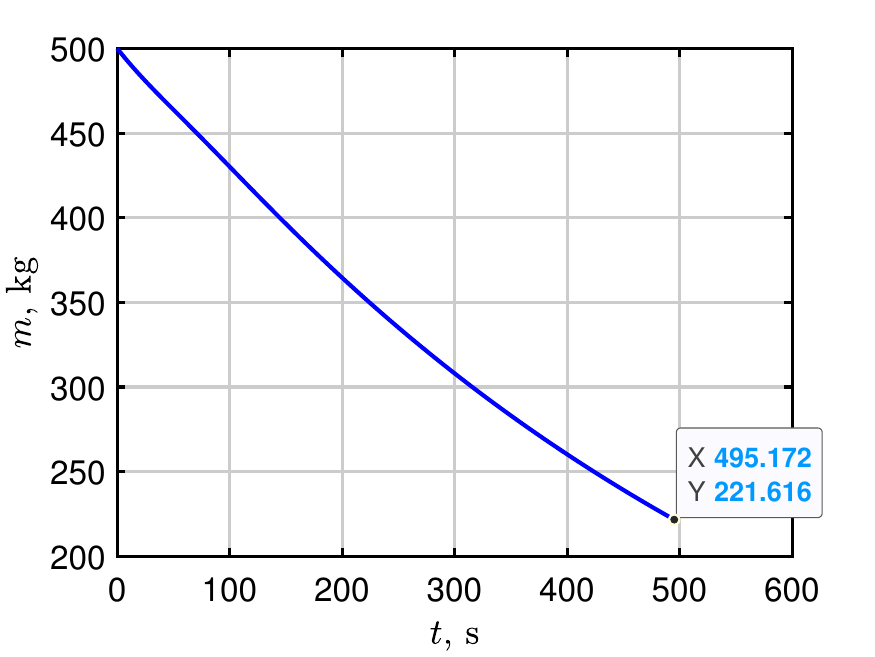}}
	\label{figk7}
	\caption{Results for case A: Sample powered descent landing }
	\label{figk}
\end{figure}

The bifurcation parameters and settling times corresponding for the PDI are given in Table~(\ref{table1}). Thus, PDT time should be chosen as $T_\textup{s}=\max \{250.4512,495.1718,341.4793 \}=495.1718$ s. Fig.~(\ref{figa}) shows the trajectory terminating at specified target with error $1.6908,10,0.077$ cm in $x$, $y$ and $z$ axes respectively, which is less than $\epsilon=0.1$ m as desired. Fig.~(\ref{figk}a) clearly shows the position histories where $z$ position reaches $5$ m directly above the landing site with $x,y$  position reducing to zero with allowed error.   Velocity monotonically reduces to zero from PDI position in Fig.~(\ref{figk}b) as shown from the analysis with maximum velocity occurring at PDI time. Acceleration components $A_\textup{x},A_\textup{y}$ initially increases then reduces to zero with monotonic nature as evident from  Fig.~(\ref{figk}c) and $A_\textup{z}$ reaching $-g$ nullifying the martian gravity. The maximum accelerations calculated analytically using Eq.~(\ref{acc}) are $(A_\textup{x}^*,A_\textup{y}^*,A_\textup{z}^*)=(0.6482,0.0769,0.6218)$ m/s$^2$, shown in Fig.~(\ref{figk}c) but, along z-axes the plot shows $A_\textup{z}^*-g=-3.0992$ m/s$^2$.  Fuel mass history in Fig.~(\ref{figk}d) shows that $221.307$ kg of fuel is consumed.
 \begin{table}[h]
	\centering
	\begin{tabular}{l  c  c c c}
		\hline \hline
		Axes & $a$ & $b\times10^{-5}$ & $c$ & $t_\textup{s}$ \\
		\hline 
		x  & -3800 &1.1080& 0  &250.4512\\
		y &  -2000&1& 0  &495.1718\\
		z & -6200 & 0.5219& 0.1616  &341.4793\\
		\hline \hline		
	\end{tabular} \\ 
	\caption{Simulation Parameters}
	\label{table1} 
\end{table}

 \subsection{Powered descent from multiple initial conditions}
\begin{figure}[h!] 
				\centering
				{\includegraphics[scale=0.55]{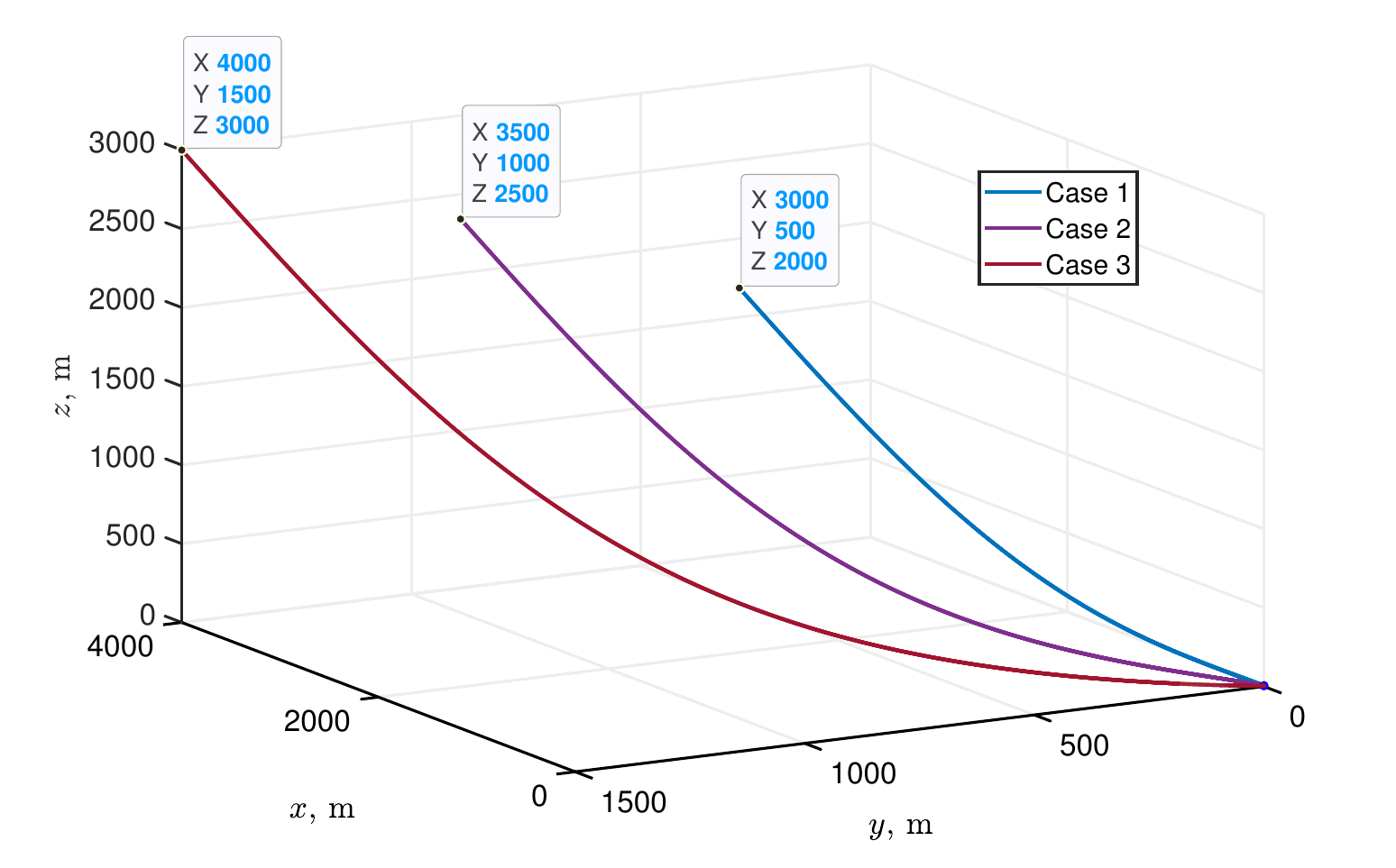}}
				\label{}
				\caption{{Powered Descent Trajectories}}
				\label{fig1s}
			\end{figure}

\begin{figure}[h!] 
	\centering
	\subfigure[{Net thrust acceleration magnitude profiles}]
	{\includegraphics[width=7.5cm]{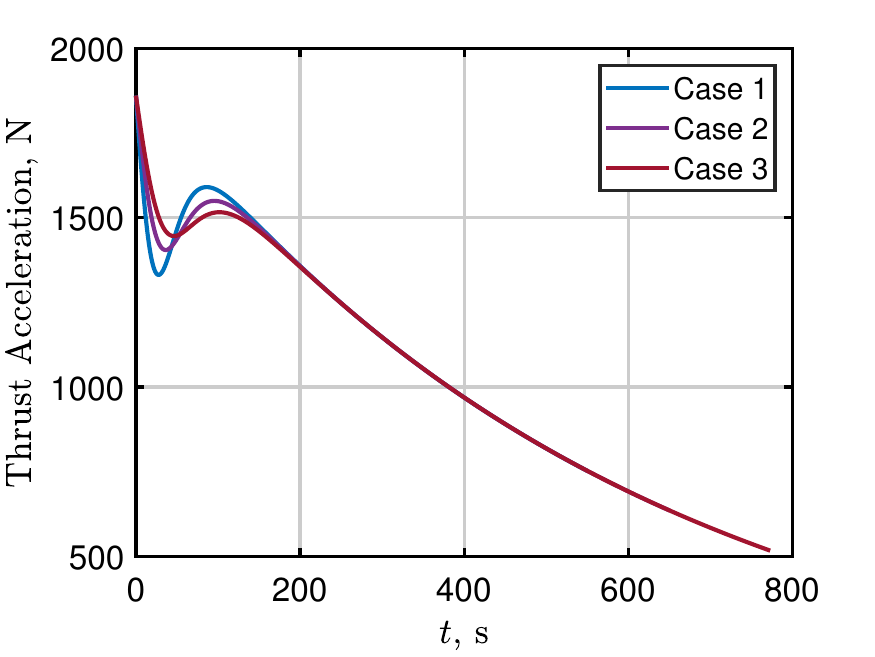}}
	\label{figk1}
	\subfigure[{Net velocity magnitude profiles}]
	{\includegraphics[width=7.5cm]{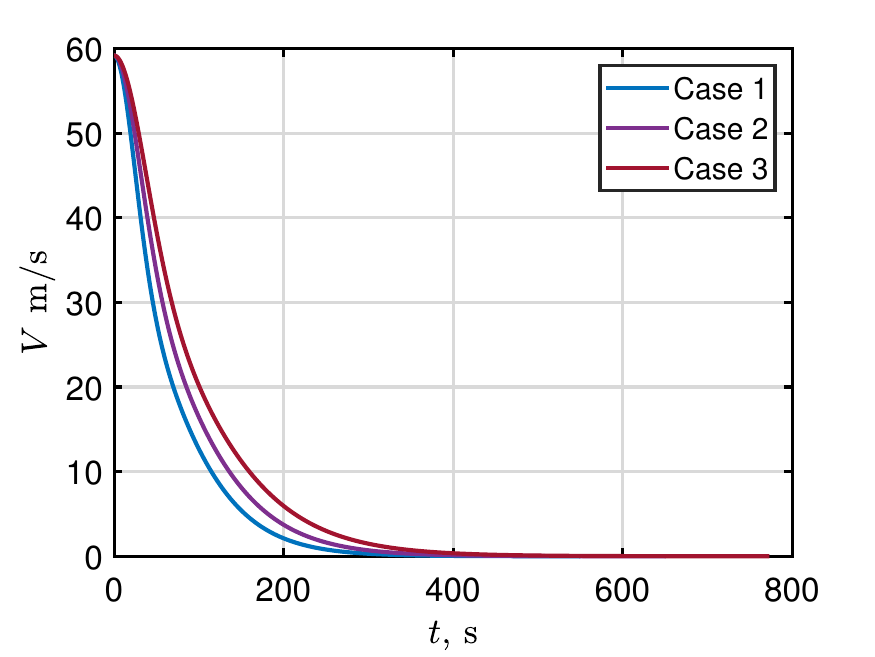}}
	\label{figk2}
 	\subfigure[{Fuel mass histories}]
	{\includegraphics[width=7.5cm]{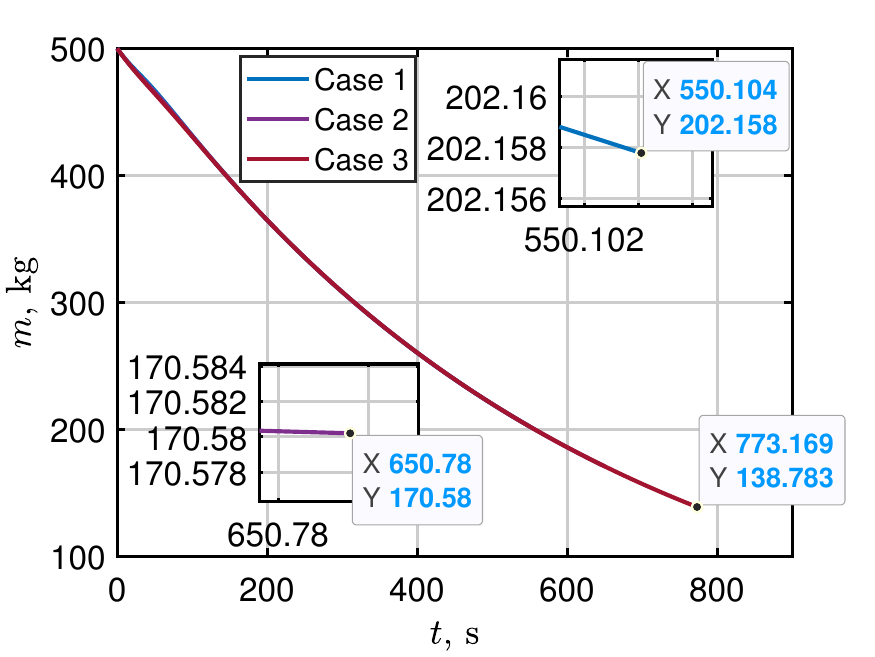}}
	\label{figk2}
	\caption{Results for case B: Multiple powered descent landings }
	\label{figk8}
\end{figure}

 This demonstration of BPDG considers different initial positions as powered descent initiation conditions. All the simulation parameters are listed in Table~(\ref{table3}). The vehicle is assumed to be heading toward the landing site with initial velocity of $(V_\textup{x0},V_\textup{y0},V_\textup{z0})=(-30,-10,-50)$ m/s. PDT condition and all other parameters are same as mentioned for Case A. Fig.~(\ref{fig1s}) shows three different powered descent trajectories for varying PDI conditions.  Fig.~(\ref{figk8}a) shows the net thrust acceleration magnitude profiles calculated as $\sqrt{\textup{T}_\textup{{x}}^2+\textup{T}_\textup{{y}}^2+\textup{T}_\textup{{z}}^2}$ N where, $\textup{T}=mA$. Net Velocity magnitude profile is shown in Fig.~(\ref{figk8}b) reducing to zero as desired. Fuel mass history is plotted in Fig.~(\ref{figk8}c) showing the amount of fuel consumed. 
  \begin{table}[h]
	\centering
	\begin{tabular}{l  c  c  c c }
		\hline \hline
		Case & $(x_0,y_0,z_0)$ km & $(a,b,c)$&  $(t_\textup{xs},t_\textup{ys},t_\textup{ys})$, s& $T_\textup{s}$, s \\
		\hline \vspace{0.5cm}
		1& $(3,0.5,2)$   & \makecell[l] {$(a_x,a_y,a_z)$=(-6000,-1000,-4000) \\  $(b_x,b_y,b_z) $ =(0.3333,4,1.2562)  $\times 10^{-5}$\\ $(c_x,c_y,c_z)$=(0,0,0.2509)} &\makecell[l] {$t_\textup{xs}=550.1041$\\ $t_\textup{ys}=230.2560$\\ $t_\textup{zs}=211.3524$} & 550.1041 \\ \vspace{0.5cm}

		2& $(3.5,1,2.5)$   & \makecell[l] {$(a_x,a_y,a_z)$=(-7000,-2000,-5000) \\  $(b_x,b_y,b_z) $ =(0.2448,1,0.8032)  $\times 10^{-5}$\\ $(c_x,c_y,c_z)$=(0,0,0.2006)} &\makecell[l] {$t_\textup{xs}=650.7804$\\ $t_\textup{ys}=495.1718$\\ $t_\textup{zs}=269.9030$} & 650.7804 \\ \vspace{0.5cm}

		3& $(4,1.5,3)$   & \makecell[l] {$(a_x,a_y,a_z)$=(-8000,-3000,-6000) \\  $(b_x,b_y,b_z) $ =(0.1875,0.4444,0.5574)  $\times 10^{-5}$\\ $(c_x,c_y,c_z)$=(0,0,0.1670)} &\makecell[l] {$t_\textup{xs}=752.6512$\\ $t_\textup{ys}=773.1689$\\ $t_\textup{zs}=329.4624$} & 773.1689 \\
		\hline \hline		
	\end{tabular} \\ 
	\caption{Simulation Parameters}
	\label{table3} 
\end{table}

\section{Conclusion} \label{sec5}
This paper presents a novel BPDG law for powered descent phase of planetary pin-point landing. The velocity dynamics of the vehicle is expressed as algebraic function of position and three bifurcation parameters. The bifurcation parameters are varied so that the stable equilibrium point is made to correspond to the guided targeting condition. Closed-form analytical expressions are derived for determining the bifurcation parameters as a function of initial position and velocity and final targeting condition. Simulation results comply with the analytic findings. Future work directions include developing the methodology considering fuel optimality and operational constraints in the BPDG problem.

\section*{Acknowledgments}
This work was carried out during the author's tenure at the Autonomous Vehicles Laboratory, Department of Aerospace Engineering, Indian Institute of Science, Bangalore, India. 

\bibliography{sample}

\end{document}